\documentclass[10pt,fleqn]{article}
\usepackage{amssymb}

\newcommand{\name}[1]{\begin{flushleft}
                       \LARGE \bf #1
                       \end{flushleft}\vspace{-3mm}}

\newcommand{\Author}[1]{\begin{flushleft}
                       \it #1 \end{flushleft}}

\newcommand{\Adress}[1]{\begin{flushleft}
                       \it #1 \end{flushleft}}

\newcommand{\be}{\begin{equation}}
\newcommand{\ee}{\end{equation}}
\newcommand{\ba}{\hspace*{-5pt}\begin{array}}
\newcommand{\ea}{\end{array}}
\newcommand{\p}{\partial}
\newcommand{\ds}{\displaystyle}

\newcommand{\pbf}[1]{\mbox{\mathversion{bold}$#1$}}

\begin{document}

\name{On the additional invariance of the Dirac and Maxwell
equations}

\medskip

\noindent{published in {\it Lettere al Nuovo Cimento},  1974,  {\bf 11}, N~10, P. 508--512.}

\Author{Wilhelm I. FUSHCHYCH\par}

\Adress{Institute of Mathematics of the National Academy of
Sciences of Ukraine, \\ 3 Tereshchenkivska  Street, 01601 Kyiv-4,
UKRAINE}

\noindent {\tt URL:
http://www.imath.kiev.ua/\~{}appmath/wif.html\\ E-mail:
symmetry@imath.kiev.ua}

\bigskip

\noindent
In this note we show that there exists a new set of operators $\{Q\}$
(this set is different from the operators which satisfy the Lie algebra
of the Poincare group $P_{1,3}$)
with respect to which the Dirac and Maxwell equations are invariant. We shall give
the detailed proof of our assertions only for the Dirac equation, since for the Maxwell
equations all the assertions are proved analogously.

The Dirac equations [1]
\be
i\frac{\p \Psi(t,\pbf{x})}{\p t}={\mathcal H}\Psi(t,\pbf{x}),
\qquad {\mathcal H}=\gamma_0 \gamma_a p_a +\gamma_0\gamma_4 m
\ee
is invariant with respect to such a set of operators $\{Q\}$
which obey the condition
\be
\left[ i\frac{\p}{\p t}-{\mathcal H}, Q\right] \Psi(t,\pbf{x})=0,
\qquad \forall \; Q\in \{Q\}.
\ee

It is well known that there are two sets of operators which satisfy the condition~(2).
The first set has the form~[2]
\be
\{\widetilde Q_1\} =\left\{\ba{l} \ds \widetilde P_0^{(1)} =p_0=i\frac{\p}{\p t},
\qquad \widetilde P_a^{(1)}=p_a =-i\frac{\p}{\p x_a}, \qquad a=1,2,3,
\vspace{2mm}\\
\ds \widetilde J_{\mu\nu}^{(1)}=x_\mu p_\nu -x_\nu p_\mu +S_{\mu\nu}, \qquad
\mu,\nu=0,1,2,3,
\ea \right.
\ee
where
\[
S_{\mu\nu}=\frac{i}{4} (\gamma_\mu\gamma_\nu -\gamma_\nu\gamma_\mu),
\qquad [x_\mu, p_\nu]=-ig_{\mu\nu}.
\]
The second set has the form~[3]
\be
\{\widetilde Q_2\}=\left\{ \ba{l}
\ds \widetilde P_0^{(2)} ={\mathcal H}=\gamma_0\gamma_a p_a +\gamma_0\gamma_4 m,
\qquad \widetilde P_a^{(2)}=p_a,
\vspace{1mm}\\
\ds \widetilde J_{ab}^{(2)}\equiv J_{ab}=x_a p_b-x_b p_a +S_{ab}, \qquad a,b=1,2,3,
\vspace{2mm}\\
\ds \widetilde J_{0a}^{(2)} =x_0 p_a -\frac 12 (x_a {\mathcal H}+{\mathcal H}x_a),
\ea \right.
\ee
We shall prove the followimg assertion.
\smallskip

\noindent
{\bf Theorem 1.} {\it The eq. (1) is invariant with respect to such two sets of operators
\be
\{\widetilde Q_3\}=\left\{ \ba{l}
\ds \widetilde P_0^{(3)} =p_0, \qquad \widetilde P_a^{(3)}=p_a,
\qquad  \widetilde J_{ab}^{(3)}\equiv \widetilde J_{ab}^{(2)}\equiv J_{ab},
\vspace{2mm}\\
\ds \widetilde J_{0a}^{(3)} =x_0 p_a -x_ap_0 -\frac i2
\left (1-\frac{\gamma_0 {\mathcal H}}{\sqrt{{\mathcal H}^2}} \right)
\left(\frac{\gamma_a}{\sqrt{{\mathcal H}^2}} -\frac{\gamma_0{\mathcal H}p_a}{{\mathcal H}^2
\sqrt{{\mathcal H}^2}}\right) p_0;
\ea \right.
\ee
\be
\{\widetilde Q_4\}=\left\{ \ba{l}
\ds \widetilde P_0^{(4)} ={\mathcal H}, \qquad \widetilde P_a^{(4)}=p_a,
\qquad  \widetilde J_{ab}^{(4)}= J_{ab}=x_a p_b-x_b p_a +S_{ab},
\vspace{2mm}\\
\ds \widetilde J_{0a}^{(4)} =x_0 p_a -\frac 12 (\widetilde x_a {\mathcal H}+{\mathcal H}
\widetilde x_a),
\ea \right.
\ee
where
\be
\widetilde x_a=x_a +\frac i2
\left (1-\frac{\gamma_0 {\mathcal H}}{\sqrt{{\mathcal H}^2}} \right)
\left(\frac{\gamma_a}{\sqrt{{\mathcal H}^2}} -\frac{\gamma_0{\mathcal H}p_a}{{\mathcal H}^2
\sqrt{{\mathcal H}^2}}\right).
\ee

}

\noindent
{\bf Proof.} It may be shown by an immediate verification that the invariant condition (2)
is satisfied for the operators (5) and (6). However, a more easy and elegant way is the
following. Let us perform a unitary transformation~[1] over eq.~(1) and the operators~(5)
and~(6)
\be
U=\frac{1}{\sqrt{2}} \left(1+\frac{\gamma_0{\mathcal H}}{\sqrt{{\mathcal H}^2}}\right).
\ee
Under the transformation eq. (1) and the operators (5), (6) will have the form
\be
\ds i \frac{\p \Phi(t,\pbf{x})}{\p t} ={\mathcal H}^c \Phi(t,\pbf{x}),
\qquad {\mathcal H}^c =\gamma_0 E, \quad \Phi=U\Psi, \qquad E=\sqrt{\pbf{p}^2+m^2},
\ee
\be
\{ Q_3\}=\left\{ \ba{l}
\ds P_0^{(3)} =U\widetilde P_0^{(3)} U^{-1} =p_0,
\qquad P_a^{(3)} =U\widetilde P_a^{(3)} U^{-1}=p_a,
\vspace{2mm}\\
J_{ab}^{(3)} =U\widetilde J_{ab}^{(3)} U^{-1} =J_{ab}, \qquad
J_{0a}^{(3)} =x_0 p_a -x_a p_0,
\ea\right.
\ee
\be
\{ Q_4\}=\left\{ \ba{l}
\ds P_0^{(4)} =U{\mathcal H} U^{-1} ={\mathcal H}^c =\gamma_0 E,
\qquad P_a^{(4)} =p_a,
\vspace{2mm}\\
\ds J_{ab}^{(4)} =U\widetilde J_{ab}^{(4)} U^{-1} =J_{ab}, \qquad
J_{0a}^{(4)} =x_0 p_a -\frac{\gamma_0}{2} (x_a E+E x_a).
\ea\right.
\ee

Now it may be readily verified that the invariant condition (2) in the new represen\-ta\-tion
\be
\left[ i\frac{\p}{\p t}-{\mathcal H}^c, Q\right] \Phi(t,\pbf{x})=0
\ee
is satisfied if the operators $\{Q\}$ have the form (10) and (11). This proves the theorem.

\smallskip

\noindent
{\bf Remark 1.} The operators (10), (11) (this means that also the operating (5), (6)) satisfy
 the relations
\be
\left[P_\mu^{(j)}, P_\nu^{(j)}\right]=0, \qquad \left[P_\mu^{(j)}, J_{\alpha \beta}^{(j)}\right]=
i\left(g_{\mu\alpha} P_\beta^{(j)} -g_{\mu\beta} P_\alpha^{(j)} \right), \qquad j=3,4.
\ee
\be
\ba{l}
\ds [J_{ab}^{(j)}, J_{cd}^{(j)}]_- =i \left(g_{cd} J_{bc}^{(j)}-g_{ac} J^{(j)}_{bd}+
g_{bc} J_{ad}^{(j)} -g_{bd} J_{ac}^{(j)}\right),
\vspace{2mm}\\
\ds \left[J_{0a}^{(j)} , J_{0b}^{(j)}\right]_-=-i\left(J_{ab}^{(j)}-S_{ab}\right),
\qquad a,b,c,d=1,2,3; \ j=3,4.
\ea
\ee
From (14) it follows that if the matrices $S_{ab}$ are added to the operators (10), (11),
then the set of operators $\left\{ P_\mu^{(j)}, S_{\mu\nu}^{(j)}, S_{ab}\right\}$
form the Lie algebra.

\smallskip

\noindent
{\bf Remark 2.} From the above considerations it follows that the wave function
$\Phi$ in passing from one inertial frame of reference to another which is moving with velocity
$-V$ may be transformed by four nonequivalent ways
\be
\ba{l}
\ds \Phi^{(j)}(t,\pbf{x})=\exp\left[i J_{0c}^{(j)} \theta_c\right]\Phi^{(j)}(t,\pbf{x}),
\qquad j=1,2,3,4,
\vspace{2mm}\\
J_{0c}^{(j)}=U\widetilde J_{0c}^{(j)} U^{-1}, \qquad \mbox{tgh}\, \theta=|V|.
\ea
\ee
It is to be emphasized that by the transformation (15) the time does not change if
$J_{0c}^{(j)}\in \{Q_2\}$ or $\{Q_4\}$:
\[
\ba{l}
x_0=\exp\left[iJ_{0c}^{(4)} \theta_c\right]x_0 \exp \left[-iJ_{0b}^{(4)}\theta_b\right]=
\exp\left[i J_{0c}^{(2)} \theta_c\right] x_0 \exp_n\left[-iJ_{0b}^{(2)}\theta_b\right]=x_0,
\vspace{2mm}\\
x_a=\exp\left[i J_{0c}^{(4)} \theta_c\right] x_a \exp \left[-iJ_{0b}^{(4)}\theta_b\right].
\ea
\]
Such transformations $x_a$, are not equivalent to the conventional Lorentz transforma\-tions.
If in these formulae $J_{0a}^{(j)} \in \{Q_3\}$, the $x_a$  and $x_0$
transform in the conventional Lorentz way. We thus find that, if the energy of a free particle
is defined as usually $E=\sqrt{\pbf{p}^2+m^2}$,
then this does not mean in general that the theory must be invariant with respect to the Lorentz
transformations.
\smallskip

\noindent
{\bf Theorem 2.} {\it The Hamiltonian ${\mathcal H}$ in eq. (1) commutes with the operators
\be
\ba{l}
\ds \widetilde S_{ab}=\frac i4 (\widetilde \gamma_a \widetilde \gamma_b -
\widetilde \gamma_b \widetilde \gamma_a), \qquad a,b=1,2,3,
\vspace{2mm}\\
\ds \widetilde S_{4a}=\frac i4 (\widetilde \gamma_4 \widetilde \gamma_a -
\widetilde \gamma_a \widetilde \gamma_4),
\ea
\ee
where
\[
\ba{l}
\ds \widetilde \gamma_a =\gamma_a +\frac 12 \left( 1-\frac{\gamma_0 {\mathcal H}}
{\sqrt{{\mathcal H}^2}} \right) \frac{(\gamma_a \gamma_c -\gamma_c \gamma_a)p_c
+2\gamma_a\gamma_4 m}{\sqrt{{\mathcal H}^2}},
\vspace{2mm}\\
\ds \widetilde \gamma_4 =\gamma_4 +\left(1-\frac{\gamma_b p_b+\gamma_4m}{\sqrt{
{\mathcal H}^2}}\right) \frac{\gamma_4 \gamma_c p_c}{\sqrt{{\mathcal H}^2}}.
\ea
\]}

\noindent
{\bf Proof.} If we perform the transformation (8) over the operators (16), we obtain
\be
S_{kl} =U\widetilde S_{kl}U^{-1} =S_{kl}=\frac i4 (\gamma_k \gamma_l -\gamma_l \gamma_k),
\qquad k,l=1,2,3,4.
\ee
From (17) it follows $[{\mathcal H}^c, S_{kl}]=0$ and
\be
[S_{kl}, S_{nr}]_-=i(g_{kr} S_{ln}-g_{kn} S_{lr} +g_{ln} S_{kr} -g_{lr} S_{kn}),
\qquad k,l,n,r=1,2,3,4.
\ee

The analogous theorem is valid for any arbitrary relativistic equation in the ca\-no\-ni\-cal form
describing free particle motion with spin~$s$~[1].

\smallskip

\noindent
{\bf Remark 3.} The operators (16) serve as an example of the nonlocal generators
(in configuration space) which satisfy the Lie algebra of the group $O_4$.
Previously it was known that the Hamiltonian had only the group $O_3$
symmetry since the spin of a particle was the integral of the motion.

Following Good~[4, 5], the Maxwell equations may be written in the Hamiltonian form
\be
\ba{l}
\ds i\frac{\p \varphi (t,\pbf{x})}{\p t} ={\mathcal H}_1 \varphi(t,\pbf{x}), \qquad
{\mathcal H}_1=\pbf{B}\pbf{p},
\vspace{2mm}\\
\ds {\mathcal H}^2\varphi\not= 0, \qquad \pbf{B}=\sigma_2 \otimes \pbf{S},
\qquad \varphi=\left(\ba{c} -\pbf{E} \\ \pbf{H} \ea \right), \qquad
\sigma_2=\left(\ba{cc} 0 & -i \\ i & 0 \ea \right).
\ea
\ee
Equations (19) by Erikson--Beckers transformation~[5]
\be
U_1=\frac{1}{\sqrt{2}} \left\{ 1+(\sigma_3\otimes 1^3)\frac{{\mathcal H}}{\sqrt{{\mathcal H}^2}}
\right\},
\qquad
1^3 =\left(\ba{ccc} 1 & 0 & 0\\ 0 & 1 & 0\\ 0 & 0& 1\ea \right).
\ee
transfer into
\be
\ba{l}
\ds i\frac{\p \Phi_1 (t,\pbf{x})}{\p t} ={\mathcal H}_1^c \Phi_1(t,\pbf{x}), \qquad
\Phi_1=U_1\varphi,
\vspace{2mm}\\
\ds {\mathcal H}_1^c= (\sigma_3\otimes 1^3)E, \qquad
\sigma_3=\left(\ba{cc} 1 & 0 \\ 0 & -1 \ea \right).
\ea
\ee
From (21) it is clear that the condition (12) (with the Hamiltonian ${\mathcal H}_1^c$)
is satisfied for $Q\in \{Q_1, Q_2\}$. Of course in (11) the $4\times 4$ matrix
$\gamma_0$ must de substituted by the matrix $\sigma_3\otimes 1^3$, and the
$4\times 4$ spin matrices by $\pbf{B}$.

\medskip
\begin{enumerate}

\footnotesize

\item Fushchych W.I., {\it Lett. Nuovo Cimento}, 1973, {\bf 6},
133,\ \ \ {\tt quant-ph/0206105};\\ Fushchych W.I., {\it Theor.
Math. Phys.}, 1971, {\bf 7}, 3 (in Russian).

\item Dirac P.A.M., The Principles of Quantum Mechanics, 4th ed., Oxford, 1958.

\item Foldy L.L., {\it Phys. Rev.}, 1956, {\bf 102}, 568.

\item Good R.H., {\it Phys. Rev.}, 1957, {\bf l05}, 1914.

\item Beckers J., {\it Nuovo Cimento}, 1965, {\bf 38}, 1362.

\end{enumerate}
\end{document}